\documentclass[prc,showpacs,twocolumn]{revtex4}

\usepackage{graphicx}
\usepackage{dcolumn}
\usepackage{bm}

\begin{document}

\newcommand{\lb}[1]{#1\!\!\!^-}

\title{Two-Step Model of Fusion for Synthesis of Superheavy Elements}

\author{Caiwan Shen$^{1,2,4,6}$, Grigori Kosenko$^{3,4}$ and Yasuhisa  Abe$^5$}

\affiliation {
         $^1$ China Institute of Atomic Energy, P.O.Box 275(18), Beijing 102413, China \\ 
         $^2$ INFN-LNS, 44 Via S. Sofia, I-95123 Catania, Italy \\
         $^3$ Department of Physics, University of Omsk, Omsk, Russia \\
         $^4$ RIKEN, 2-1 Hirosawa, Wako-shi, Saitama 351-0198, Japan \\
         $^5$ Yukawa Institute for Theoretical Physics, Kyoto University, Kyoto, Japan \\
         $^6$ Center of Theoretical Nuclear Physics, National Laboratory of Lanzhou, 
              Lanzhou, China
        }


\begin{abstract}
A new model is proposed for fusion mechanisms of massive nuclear systems where
so-called fusion hindrance exists.  
The model describes two-body collision processes in an approaching
phase and shape evolutions of an amalgamated system into the compound 
nucleus formation. 
It is applied to $^{48}$Ca-induced reactions and is found to
reproduce the experimental fusion cross sections extremely
well, without any free parameter.
Combined with the statistical decay theory, residue cross sections for 
the superheavy elements can be readily calculated.  Examples are given.
\end{abstract}

\pacs{24.60.-k, 25.60.Pj, 25.70.-z, 25.70.Jj}

\maketitle


How many elements exist in the nature or what is the 
heaviest element has been an intriguing question since the periodic table was proposed 
for the chemical elements.  The heaviest element that exists in the nature is now known 
to be Uranium with atomic number Z being 92.  But the discovery of the magic 
numbers in atomic nuclei and their understanding by the shells of nucleonic motion[1]  
suggest that much heavier atomic nuclei might exist, stabilized by
extra-bindings  
due to possible shells next to the largests known, i.e., Z=82 and N=126.    Actually, 
many theoretical calculations have been made, predicting the next double closed shell 
nucleus to be with Z=114, 120, or 126 and N=184[2].  Naturally, enormous 
experimental efforts have been devoted to finding out traces of
existence of the corresponding 
superheavy atomic nuclei and to synthesizing them with nuclear reactions, especially 
with heavy-ion fusion reactions [3].  But what combination of ions is
favorable as  
entrance channels and what incident energy is the optimum for residues are not 
predicted well, and thus, the experiments have been performed
according to the results of systematic studies done so far. 
This is due to the lack of our knowledge of reaction mechanisms.  

Based on the theory 
of compound nucleus reactions, the residue cross sections are given as follows,
\begin{equation}
\sigma_{\mbox{\scriptsize res}}=\pi{\lb{\lambda}}^2\Sigma_J(2J+1)\cdot
P^J_{\mbox{\scriptsize fusion}}(E_{\mbox{\scriptsize c.m.}})\cdot P^J_{\mbox{\scriptsize surv}}(E^*),
\end{equation}
\noindent
where $\lb\lambda$ is the inverse of the wave number and $J$ is the total angular 
momentum quantum number.    $P_{\mbox{\scriptsize fusion}}$ and $P_{\mbox{\scriptsize surv}}$ denote the fusion and the survival 
probabilities, respectively.  The latter is given by the statistical theory of decay, i.e., by 
competitions between neutron emission and fission decay.  
Essentially unknown is 
the fusion probability, i.e., fusion mechanism of massive systems,
although there are 
ambiguities in the parameters in the properties of heavy and superheavy nuclei which 
give rise to uncertainties in calculating the survival probability. 

In lighter systems, 
the fusion probability is well determined by the barrier defined with the Coulomb and 
the nuclear attraction between nuclei in the entrance channel, but in
massive systems, the situation is not so simple.  It has been well
known 
experimentally that there is the fusion-hindrance [4], which is often described with 
so-called extra-push energy which is required for a system to fuse in addition to the 
barrier height [5].  A physical origin or mechanism is not yet well clarified.  There are 
two possible interpretations proposed.  They both attribute it to energy dissipations; 
one is due to the dissipation of the initial kinetic energy during two-body collisions 
passing over the barrier [6], while the other is due to the dissipation of the energy of  
collective motions which would lead an amalgamated system to the spherical 
compound nucleus [5].  It is natural to consider that both mechanisms exit, though we 
don't know a priori which one dominates in which situation.  We, thus, propose a new 
theoretical framework for fusion, i.e., a two-step model which incorporates both of them 
properly [7].  

In the approaching phase of passing over the Coulomb barrier, we describe the 
system as collision processes under frictional forces, up to the contact point of two 
incident nuclear matters and then, we describe dynamical evolutions of the 
amalgamated mono-nuclear system toward the spherical shape under frictional forces 
acting in collective motions of excited nuclei.  As is given below, both dynamical 
processes are described by Langevin equations which include random forces 
associated to the respective frictions.   It would be worth to mention here that the 
fluctuations due to the random forces are crucially important in problems of small 
probability such as in syntheses of the superheavy elements (SHE), because we have 
to investigate cases where mean trajectories never reach the spherical
shape.  Another point to be mentioned is that since 
the two steps are connected successively, the results of the first
step not only gives a probability for incident 
ions to stick to each other (sticking probability
$P_{\mbox{\scriptsize stick}}$) but also gives 
initial conditions for the second step.  
Thus, the method of the connection from the first to the second steps
is natural, which is neither related to the diabaticity nor the
adiabaticity.  It is completely new and could be called as
``statistical'', as will be seen below.
In massive systems, there is a 
conditional saddle point, or a ridge line between the amalgamated configuration and the 
spherical shape on the potential energy surface calculated with the liquid drop model 
(LDM), which could be considered to be another barrier inside and makes most 
trajectories to return back to re-separation (quasi-fission, etc.),
i.e., gives 
rise to a small probability for forming the spherical shape (formation probability $P_{\mbox{\scriptsize form}}$).  
Thus, the fusion probability is given by the product of the two 
probabilities,  
\begin{equation}
P^J_{\mbox{\scriptsize fusion}}(E_{\mbox{\scriptsize c.m.}})
=P^J_{\mbox{\scriptsize stick}}\left(E_{\mbox{\scriptsize c.m.}})
\cdot P^J_{\mbox{\scriptsize form}}(E_{\mbox{\scriptsize c.m.}}\right).
\end{equation}
In order to realize the model, we employ the surface friction model
(SFM)[8] for 
the approaching phase and the one-body wall-and-window formula[9] of the dissipation 
for the shape evolutions, i.e., for the second step.

As for the approaching phase, the equation of motion is only for the 
radial degree of freedom and the orbital angular momentum, and is given below,
\begin{eqnarray}\nonumber
&&\frac{dr}{dt}=\frac{1}{\mu}p,\\\nonumber
&&\frac{dp}{dt}=-\frac{dV}{dr}-\frac{\partial}{\partial
r}\frac{\hbar^2L(t)^2}{2\mu r^2}-C_r(r)\frac{p}{\mu}+R_r(t),\\
&&\frac{dL(t)}{dt}=-\frac{C_L(r)}{\mu}\cdot(L(t)-L_{st})+R_T(t),
\end{eqnarray}
\noindent
where $\mu$ is the reduced mass of the collision system, and $V$ is
the sum of the 
Coulomb potential $V_c$ and the nuclear potential $V_n$.   $C_i(r)$
is the radial or the tangential 
friction coefficient which is assumed to have the following form factor,
\begin{equation}
C_i(r)=K^0_i\cdot\left({dV_n}/{dr}\right)^2,
\end{equation}
\noindent
where $K^0_r=0.035$ and $K^0_T=0.0001$ in unit of 10$^{-21}$s/MeV.
$R_i$ denotes a random force associated with the friction for $i=r$
(radial) or $T$ (tangential), and assumed to be 
Gaussian, and to satisfy the following property,
\begin{eqnarray}\nonumber
&&\langle R_i(t)\rangle=0,\\
&&\langle R_i(t)\cdot
R_j(t')\rangle=2\delta_{ij}\delta(t-t')\cdot C_i\left(r(t)\right)T^J(t), 
\end{eqnarray}
\begin{figure}[hbt]
\includegraphics[width=7.2cm]{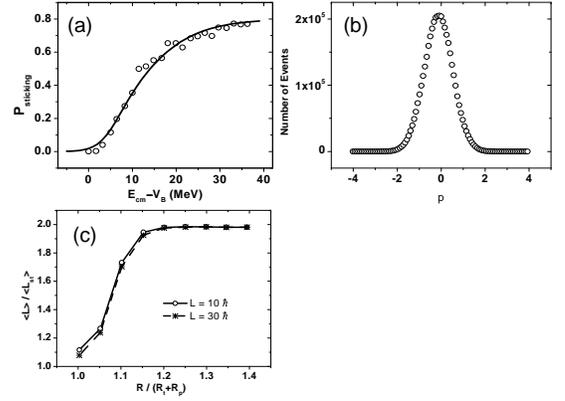}
\caption{Results on $^{48}$Ca-$^{238}$U system with SFM.  The sticking
probability for L=0 is shown in (a) (curve is for an eye-guide), the
radial momentum distribution
in (b) in unit of 10$^{-21}$sec$\cdot$MeV/fm and the average orbital angular momentum v.s. the relative
distance in (c).}
\end{figure}

\noindent
where the last equation is the fluctuation-dissipation theorem with
temperature $T^J(t)$, $J$ being equal to a total angular momentum of
the system, i.e., an incident orbital angular momentum $L$.    
$L_{\mbox{\scriptsize st}}$ denotes the limiting orbital angular
momentum under the friction, which is so-called the sliding limit in
the SFM and is equal to $5/7\cdot L$.
We calculate many 
trajectories over relevant impact parameters and obtain probabilities for their reaching 
the contact point, respectively.  Fig.1a shows the calculated sticking probability for L=0 
for the case of $^{48}$Ca-$^{238}$U system.  Incident energy is given relative to the barrier 
height.   It is readily seen that at energies just above the barrier there is almost no 
probability.   This is due to the fact that the form factor assumed in SFM stretches 
over outside the barrier top position in massive systems.   The results already appear 
to explain the fusion hindrance and at least partially the extra-push
energy, while the second step is also expected to give rise to an
additional contribution.   In order to know the physical situation at 
the contact point, we analyze the radial momentum distribution as well as that of the 
orbital angular momentum.   The radial momentum distribution is found 
to be almost purely Gaussian, as shown in Fig.1b.  Its width is in consistence with the 
temperature of the heat bath of nucleons which is supposed to absorb
the initial
kinetic energy through the friction force.   The example shown is for
$L=0$, but the other 
angular momentum cases behave in the same way.   Therefore, the calculated 
distribution $S^J(p_0,E_{\mbox{\scriptsize c.m.}})$ can be expressed
as follows for each angular momentum,
\begin{equation}
S^J(p_0, E_{\mbox{\scriptsize c.m.}})=P^J_{\mbox{\scriptsize
stick}}(E_{\mbox{\scriptsize c.m.}})\cdot g^J(p_0, \bar p^J_0, T^J_0),
\end{equation}
\noindent%
where the normalized Gaussian distribution \break
$g^J(p_0, \, \bar p^J_0, \, T^J_0)$
is given generally so as to include an 
average mean momentum left $(\bar p^J_0)$  which is almost equal to zero in the present case.   This 
distribution is used as the initial inputs to the dynamical evolutions
in the second step, i.e., to Eq.~(8) below.   
$T^J_0$ denotes the temperature of the amalgamated system.
The total energy available for the compound nucleus $E^*$ is written
by the energy conservation as follows,
\begin{equation}
E^*=E_{\mbox{\scriptsize c.m.}}+Q=V_0-E_{\mbox{\scriptsize shell}}+\varepsilon_0+k_0,
\end{equation}
\noindent
where $Q$ denotes the $Q$-value of the fusion reaction.  
$E_{\mbox{\scriptsize shell}}$ the shell correction energy of the
ground state, $V_0$ the LDM
potential energy of the contact point, $\varepsilon_0$ the intrinsic
excitation, and $k_0$ the radial kinetic energy left
at the contact point.   
The latter two are averagely given as 
$a_0\cdot {T^J_0}^2$ and $(\bar{p}^{J}_0)^{2}/2\mu+\frac12T^J_0$,
respectively with the level density parameter $a_0$ which is
calculated according to T\"oke and Swiatecki [10].
The orbital 
angular momentum is also analyzed.   The average value is plotted as a function of the 
radial distance in Fig.1c.   It is seen that it approaches to the dissipation limit $L_{st}$ 
about the contact point, which indicates that the incident system reaches the sticking 
limit, if the rolling friction is properly taken into account.    We, thus, can consider 
that the relative motion is completely damped and reaches the thermal equilibrium 
with the heat bath at the contact point, i.e., that the incident ions form an amalgamated 
mono-nuclear system, the probability of which depends on the incident
energy and is extremely small just above the barrier.
It should be noticed here that $\bar p^J_0=0$ does not always hold, for
example not in $^{100}$Mo-$^{100}$Mo system etc.

Subsequent shape evolutions of the pear-shaped 
mono-nucleus formed with the incident ions are described by the multi-dimensional 
Langevin equation which is the same as that used for dynamical studies of fission[11],
\begin{eqnarray}\nonumber
&&\frac{dq_i}{dt}=(m^{-1})_{ij}\cdot p_j,\\\nonumber
&&\frac{dp_i}{dt}=-\frac{\partial U^J}{\partial
q_i}-\frac12\frac{\partial}{\partial q_i}(m^{-1})_{jk}\cdot p_j\cdot
p_k,\\\nonumber
&&\hspace{1.1cm}-\gamma_{ij}\cdot(m^{-1})_{jk}\cdot p_k+g_{ij}\cdot R_j(t),\\
&&g_{ik}g_{jk}=\gamma_{ij}\cdot T^J,
\end{eqnarray}
\begin{figure}[htb]
\includegraphics[width=5.2cm,height=4.cm]{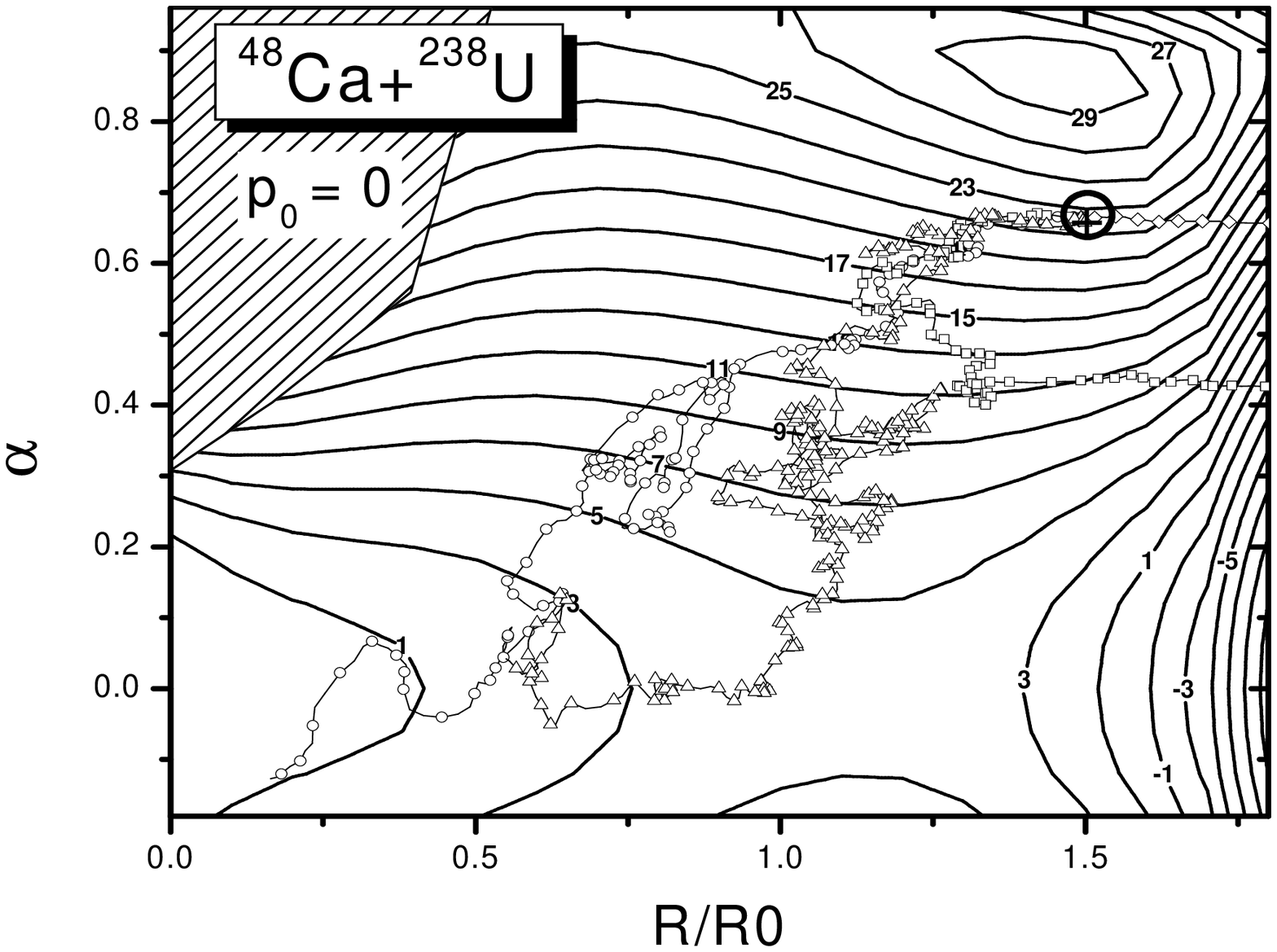}
\caption{Examples of the trajectories are displayed with the same
initial radial momentum being equal to zero.  Random force gives rise
to a variety of the trajectories.  The circle in the upper right corner
corresponds to the touching configuration reached by the first step,
from which dynamical evolutions of shape start.  ($R_0$ being the
radius of the spherical ground state) }
\end{figure}

\noindent%
where summation is implicitly assumed over repeated suffixes.  
The collective mass tensor $m_{ij}$ is the hydrodynamical one
and the potential $U^J$ is
calculated by the finite range LDM with two-center parameterization of nuclear 
shapes[12],   
added with 
the rotational energy of the system calculated
with the rigid body moment of inertia.
The random force $R_i(t)$ is again gaussian with the normalization 2,
and the tensor $g_{ij}$ is now 
related to the friction tensor $\gamma_{ij}$, as is given in the last equation, i.e., the 
generalized Fluctuation-Dissipation theorem in the multidimensional case.    The 
friction tensor is calculated with the wall-and-window formula [9].   The temperature 
$T^J$ of the heat bath is better to be taken to be that at the conditional saddle point, but is 
approximated with that at the contact point, i.e. $T^J_0$.   They are close to each other for 
the $^{48}$Ca-induced reactions.    In the present calculations we only use the relative 
distance $R$ and the mass asymmetry coordinate $\alpha$ with the other degrees
of freedom being 
frozen.   For example, the neck parameter is taken to be 0.8, based on our 
experiences that it does not change so much during passing over the conditional saddle 
point in the three-dimensional calculations.   
Fig.2 
shows examples of the trajectories on the LDM potential for $^{48}$Ca-$^{238}$U system for 
initial radial momenta and thus initial energies being equal to
zero. Calculations of many trajectories, starting with 
various initial radial momenta give a distribution of formation
probability $F^J(p_0, T^J)$.    
By making a convolution of it with the Gaussion distribution of the initial momentum 
$g^J(p_0,\bar p^J_0,T^J_0)$, we obtain the formation probability
$P_{\mbox{\scriptsize form}}$
\begin{equation}
P_{\mbox{\scriptsize form}}(E_{\mbox{\scriptsize c.m.}})=\int
dp_0F^J(p_0, T^J)\cdot g^J(p_0, \bar p^J_0, T^J_0).
\end{equation}
\begin{figure}[htb]
\includegraphics[width=3.5cm]{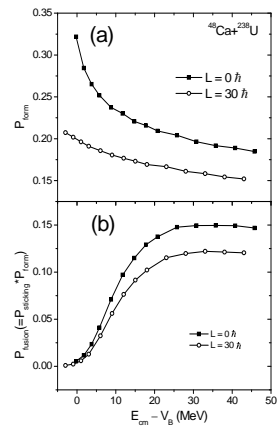}
\caption{Calculated formation and fusion probabilities are shown in (a) 
and (b), respectively.}
\end{figure}
Fig.3a shows the calculated formation probability for $^{48}$Ca-$^{238}$U system for L=0 and 
30.   
In Fig.3b, the final fusion 
probability is plotted versus incident energy.      
At the first glance, the decreasing energy dependences seem to be
peculiar, but the energy dependence of the passing-over probability
under friction delicately depends on the strength of friction and the
incident momentum.
Actually, slightly weaker friction gives rise to an increasing energy
dependence.  
A detailed analysis with the 1-dimensional model will be given
elsewhere [15]. 
It should be also mentioned here that the 
present model is completely classical, and thus there is no quantum
tunneling effect 
included, which limits the lowest energy to be reached.

Fusion cross sections are 
calculated with the fusion probability as usual,                     
$\sigma_{\mbox{\scriptsize
fusion}}=\pi{\lb{\lambda}}^2\Sigma_J(2J+1)\cdot P^J_{\mbox{\scriptsize
fusion}}(E_{\mbox{\scriptsize c.m.}})$,
and are shown in 
Fig.4 for the four systems with $^{48}$Ca beam, together with some measured 
cross sections[14].   It is extremely surprising that the calculations
well reproduce the 
experiments without any adjustment of
the model parameters. 
Experimental measurements are highly desirable in other heavy systems for 
comparisons with the present calculations.
\begin{figure}[t]
\includegraphics[width=6.2cm]{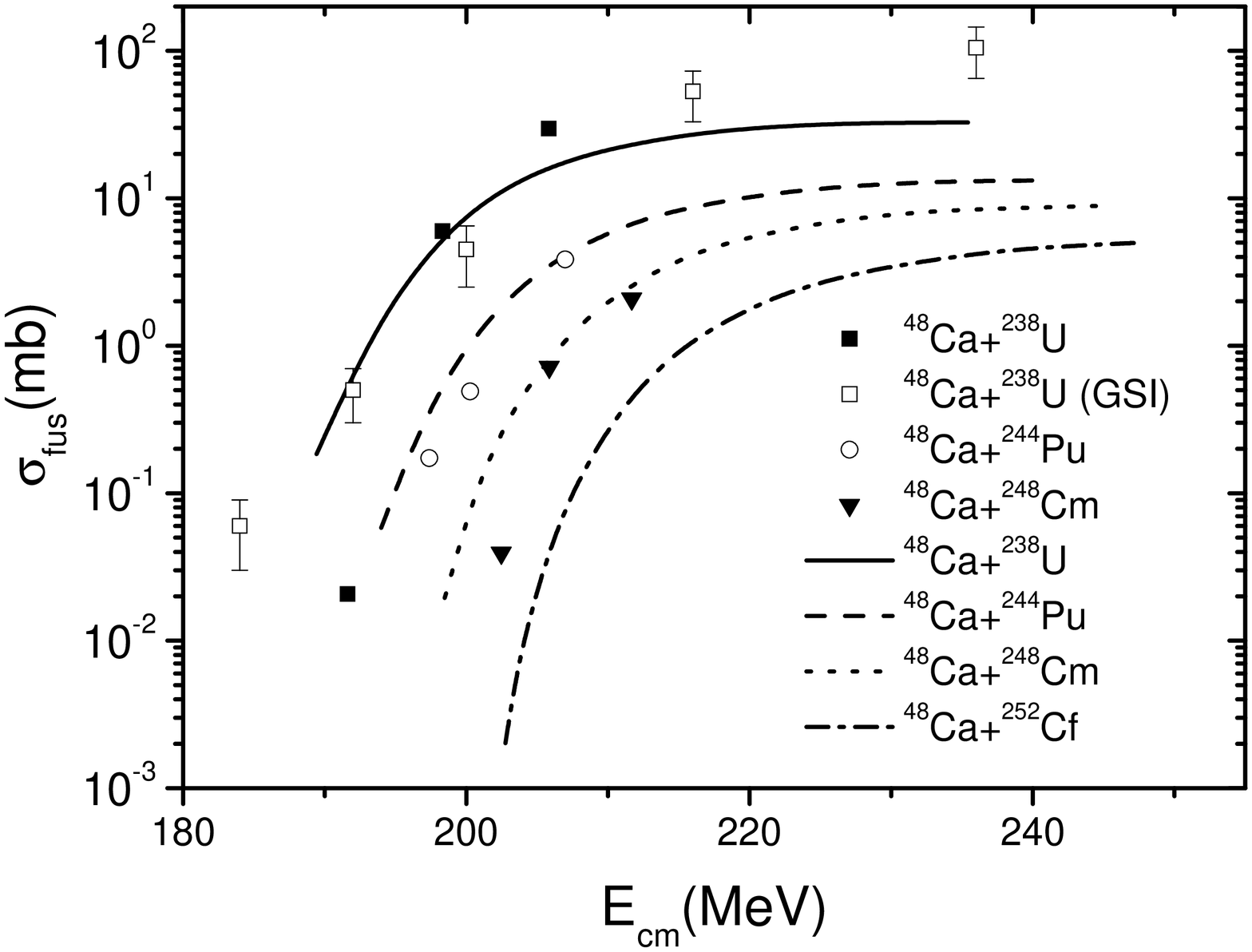}
\caption{Calculated excitation functions of fusion reactions for
$^{48}$Ca-$^{244}$Pu, -$^{248}$Cm and -$^{252}$Cf systems, together with
the available experimental data [13] for GSI and [14] for Dubna.}
\vspace{-3mm}
\end{figure}

\begin{table}[t]
\caption{Calculated maximum residue cross sections of the three
systems $^{48}$Ca-$^{244}$Pu, -$^{248}$Cm and -$^{252}$Cf are summarized.
The factor 1/3 for M\"oller masses is rather arbitrarily chosen.  
($\sigma _{\max }$:pb, $E^{\ast }$: MeV)}
\begin{tabular}{c|lc|cc|cc}
\hline\hline
& \multicolumn{2}{|c|}{Prediction of} & \multicolumn{2}{|c|}{3n} & 
\multicolumn{2}{|c}{4n} \\ \cline{4-7}
$^{48}$Ca & \multicolumn{2}{|c|}{$\Delta E_{shell}$(MeV)} & $\sigma
_{\max }$ & $E^{\ast }$ & $\sigma _{\max }$ & $E^{\ast }$ \\ \hline
$^{244}$Pu & Liran & $-0.23$ & 0.018 & 30.6 & 0.018 & 36.5 \\ 
& M\"oller/3 & $-2.96$ & 7.39 & 30.1 & 6.00 & 35.3 \\ 
& \multicolumn{2}{|l|}{Experiment} & $\approx 1$ & $E_{\text{lab}}$=236 & $%
\approx 1$ & $E_{\text{lab}}$=236 \\ \hline
$^{248}$Cm & Liran & $-1.37$ & 0.254 & 31.1 & 0.045 & 37.8 \\ 
& M\"oller/3 & $-2.86$ & 4.56 & 30.4 & 2.98 & 35.6 \\ 
& \multicolumn{2}{|l|}{Experiment} &  &   & $0.6$ & 35.8 \\  \hline
$^{252}$Cf & Liran & $-3.24$ & 1.057 & 32.7 & 0.095 & 38.2 \\ 
& M\"oller/3 & $-2.41$ & 0.216 & 28.8 & 0.086 & 33.5 \\ \hline\hline
\end{tabular}
\end{table}

In order to show that we are ready for calculations of residue cross
 sections for SHE, we give examples for Z=114, 116 and 118, by the use 
 of $P_{\mbox{\scriptsize surv}}$ calculated with HIVAP [16].
Actually, the shell correction energies are the most
crucial quantities in 
residue calculations, because they effectively give the fission
barriers for SHEs. 
And they are not yet firmly predicted, and thus we take those by
 M\"oller and Liran [17] as typical examples of 
mass predictions, and compare 
with the recent Dubna experiments [18], which are given in Table I.

In brief, the new two-step model has been found to be extremely successful in reproducing 
the available fusion data of $^{48}$Ca induced reactions.    
By combining 
the present fusion probabilities with the standard statistical decay calculations, we 
have obtained residue cross sections for Z=114, 116, and 118, which
are in a reasonable 
agreement with the recent Dubna experiments, but with rather small shell correction 
energies, much smaller than previously thought.
A systematic study of residue cross sections are being made.
Furthermore, the model is now being applied 
to other massive systems, such as $^{100}$Mo-$^{100}$Mo etc.   

This work is partly supported by Chinese Academy of Science Knowledge
Innovation Project (KJCX1-N11) and by the Grant-in-Aids of JSPS (no.~13640278).

\end{document}